# Quantitative investigation of low-dose PET imaging and post-reconstruction smoothing

Behnoush Sanaei[1], Reza Faghihi[1] and Hossein Arabi[2]

[1] Division of Medical Radiation, Department of Nuclear Engineering, Shiraz University, Shiraz, Iran

[2] Division of Nuclear Medicine and Molecular Imaging, Department of Medical Imaging, Geneva University Hospital, CH-1211 Geneva 4, Switzerland

**Abstract**

Positron emission tomography (PET) imaging is commonly used in clinical practices for cancer diagnosis, tumor detection, evaluating lesion malignancy, and the staging of diseases. In PET imaging, a high amount of radiotracer should be injected to form high-quality PET images; which will cause a high risk of radiation damage to patients. On the other hand, reducing the injected dose to avoid potential radiation risks would lead to poor image quality as well as noise-induced quantitative bias. In this study, $^{18}$F-FDG PET/CT brain scans of 50 patients with head and neck malignant lesions were employed to systematically assess the relationship between the amount of injected dose (10%, 8%, 6%, 5%, 4%, 3%, 2%, and 1% of standard dose) and the image quality through measuring standard image quality metrics (peak-signal-to-noise-ration (PSNR), structural similarity index (SSIM), root mean square error (RMSE), and standard uptake value (SUV) bias) for the whole head region as well as within the malignant lesions, considering the standard-dose PET images as reference. Furthermore, we evaluated the impact of post-reconstruction Gaussian filtering on the PET images in order to reduce noise and improve the signal-to-noise ratio at different low-dose levels. Significant degradation of PET image quality and tumor detectability was observed with a decrease in the injected dose by more than 5%, leading to a remarkable increase in RMSE from 0.173 SUV (at 5%) to 1.454 SUV (at 1%). The quantitative investigation of the malignant lesions demonstrated that $SUV_{max}$ bias greatly increased in low-dose PET images (in particular at 1%, 2%, 3% levels) before applying the post-reconstruction filter, while applying the Gaussian filter on low-dose PET images led to a significant reduction in $SUV_{max}$ bias. The $SUV_{mean}$ bias within the malignant lesions was negligible (less than 1%) in low-dose PET images; however, this bias increased significantly after applying the post-reconstruction filter. In conclusion, it is strongly recommended that the $SUV_{max}$ bias and $SUV_{mean}$ bias in low-dose PET images should be considered prior to and following the application of the post-reconstruction Guassain filter, respectively.

## 1. Introduction

Positron emission tomography (PET) imaging is frequently used in clinical practices for purposes such as cancer diagnosis, tumor detection, evaluating lesion malignancy, and the staging of diseases [1, 2]. In order to acquire high-quality PET images, a standard dose of radioactive tracer is injected into the patient prior to the image acquisition, then the PET system detects pairs of gamma rays emitted indirectly from the radiotracer concentration within the body. The map of radioactive uptake within the body is then obtained through the reconstruction of the recorded signals [3-6].

Although using a large amount of radioactive tracer would lead to a better quality of PET images (owing to the reduced statistical noise and the improved signal-to-noise ratio [SNR]), it will inevitably cause a high risk of radiation damage to patients and healthcare providers [7-9]. On the other hand, since noise levels in the PET images highly depend on the amount of injected radiotracer as well as the duration of the image acquisition [10], reducing the injected dose in order to avoid the radiation risk would lead to noise-induced quantitative bias and image artifacts which may adversely affect the clinical value of PET images [11, 12].

A number of studies have been carried out to investigate the impact of reducing the injected dose on PET image quality. Schaefferkoetter et al. [13] assessed different levels of low-dose PET imaging using nine ranges of the total registered counts from 250K to 40M. It was demonstrated that the PET images reconstructed with 5 million or greater counts represent a relatively high level of lesion detectability with only slight differences. Yan et al. [14] investigated the relationship between the levels of noise, SNR, and contrast to noise ratio (CNR), and the injected dose or the number of counts to determine the clinically tolerable (acceptable quality of PET images) reduced dose levels in PET imaging. Similarly, Oehmigen et al. [15] evaluated the quality of PET/MR images as the function of injected radiotracer activity and acquisition duration using a standardized phantom. The abovementioned studies focused on the impact of injected dose on PET image quality, however, they did not consider a systematic dose reduction to establish an optimal compromise between the reduced dose (patient absorbed dose) and the image quality (clinical value of the PET images) [16, 17].

In this study, we set out to evaluate the relationship between the amount of injected dose in brain PET imaging and the image quality through measuring the standard image quality metrics such as peak-signal-to-noise-ration (PSNR), structural similarity index (SSIM), root mean square error (RMSE), and standard uptake value (SUV) bias considering the standard dose PET images (acquired with standard injected dose) as reference. The image quality metrics and quantitative bias were calculated for the entire head region as well as the abnormally high-uptake regions in the head and neck. To this end, only patients with malignant lesions in the head and neck were included in this study. Moreover, the impact of applying conventional post-reconstruction smoothing on PET images in order to reduce the noise (or recovering the image quality) was assessed with a particular focus on the malignant lesions.

## 2. Materials and Methods

This work deals with the generation of low-dose PET images at different levels from standard dose data to be compared against reference images (standard dose data) using standard metrics for the assessment of the image quality. Then, a post-reconstruction Gaussian filter was applied to the low-dose PET images to investigate the overall image quality improvement and in particular the quantification accuracy within the malignant lesions.

### 2.1. Image acquisition and low-dose PET generation

The study was conducted on 50 $^{18}$F-FDG PET/CT brain scans from patients with head and neck malignant lesions (24 males and 26 females, 71 ± 9 yrs, mean age ± standard deviation (SD)). The PET scans were performed for an acquisition time of 20-min after the injection of 210 ± 8 MBq of $^{18}$F-FDG. The PET/CT acquisitions were performed on a Biograph-6 scanner (Siemens Healthcare, Erlangen, Germany) about 40 minutes after the injection. The PET raw data, recorded in list-mod format, was randomly sampled to contain 10%, 8%, 6%, 5%, 4%, 3%, 2%, and 1% of the total counts in the standard/high-dose PET imaging. In this regard, the low dose versions of the standard PET images with the abovementioned percentages were reconstructed using ordered subsets-expectation maximization (OSEM) algorithm (4 iterations, 18 subsets). No post-reconstruction filtering was applied to the low-dose PET images.

### 2.2 Optimization of post-reconstruction smoothing

In order to reduce the adverse impact of noise on the PET images, post-reconstruction Gaussian filtering is usually employed in clinical practices [18]. The Gaussian filter kernel (level of smoothness) is normally determined according to the levels of noise in the PET images [19]. In this light, to reduce noise levels in the different low-dose PET images (namely 10%, 8%, 6%, 5%, 4%, 3%, 2%, and 1%), the Gaussian filter kernels should be specifically optimized for each noise level to reach the highest signal to noise ratio (SNR) at each reduced dose level. To obtain the optimal Gaussian kernel (or standard deviation of the Gaussian filter) at each reduced dose level, different values were assessed for the Gaussian kernel. The kernels which led to the maximum peak to noise ratio (PSNR) at each reduced dose level were selected as the optimal kernels [20]. To measure the PSNR metric at each reduced dose-level, the full-dose/standard PET image was considered as the reference. The quality of low-dose PET images before and after applying the optimal Gaussian filters was separately evaluated at each reduced dose level for the entire head region as well as within the malignant lesions.

## 2.3 Evaluation strategy

*Whole-brain analysis:* The image quality of low-dose PET images within the entire head region was evaluated using three standard metrics including peak-to-signal-noise-ratio (PSNR), structural similarity index metric (SSIM), and root mean square error (RMSE).

The peak signal to noise ratio (PSNR) was calculated as the ratio between the maximum value of signals in either the standard-dose (S-PET) or low-dose (L-PET) PET images and the mean square error between S-PET and L-PET images (Eqs. 1 and 2).

$$MSE = \frac{\sum_{M,N}(S-PET(M,N)-L-PET(M,N))^2}{M*N} \quad (1)$$

$$PSNR = 10\ \log_{10}(\frac{R^2}{MSE}) \quad (2)$$

where $M$ and $N$ denote the number of rows and columns in the input images (S-PET and L-PET), and $R$ is the maximum value of the input images.

The structural similarity index metric (SSIM) was used to describe the similarity between S-PET and L-PET images using Eq. 3.

$$SSIM(S,L) = \frac{(2\mu_S\mu_L+c_1)(2\sigma_{SL}+c_2)}{(\mu_S^2+\mu_L^2+c_1)(\sigma_S^2+\sigma_L^2+c_2)} \quad (3)$$

$$c_1 = (0.01d)^2 \qquad c_2 = (0.03d)^2$$

Where $\mu_S$, $\mu_L$, and $\sigma_S^2$, $\sigma_L^2$ represent the average and the variance of S-PET and L-PET images, respectively, $\sigma_{SL}$ is the covariance of the S-PET and L-PET images, and $d$ is the dynamic range of the pixel values.

The root mean square error (RMSE), describing the difference between the standard uptake value (SUV) in the S-PET image and L-PET images was calculated using Eq. 4.

$$RMSE = \sqrt{\frac{1}{N}\sum_N(SUVs_i - SUVl_i)^2} \quad (4)$$

Where $SUVs_i$ and $SUVl_i$ are SUV at voxel $i$ for the S-PET image and L-PET images, respectively, and $N$ is the total number of voxels.

*Lesion analysis*

The quality of low-dose PET images before and after applying the post-reconstruction filter was evaluated for the entire head regions using the above-mentioned metrics. In addition, the quantitative assessment of the low-dose PET imaging was performed for the malignant lesions in the head and neck region. The accuracy of activity concentration within the fifty manually defined tumors on the reference S-PET images was assessed using the RMSE, $SUV_{max}$ bias, and $SUV_{mean}$ bias metrics.

$SUV_{max}$ bias was calculated for the lesions on the L-PET images with respect to reference S-PET images using Eq. 5 in which maximum SUV represents the pixel with the highest radiotracer uptake.

$$SUV_{max}\ bias(\%) = 100 \times \frac{SUV_{max,S} - SUV_{max,L}}{SUV_{max,S}} \quad (5)$$

Where $SUV_{max,S}$ and $SUV_{max,L}$ are the maximum standard uptake value within the lesion in the S-PET and L-PET images, respectively.

$SUV_{mean}$ bias was also calculated using Eq. 6 wherein $SUV_{mean,S}$ and $SUV_{mean,L}$ denote mean standard uptake value in the lesion for the S-PET and L-PET images, respectively.

$$SUV_{mean}\ bias(\%) = 100 \times \frac{SUV_{mean,S} - SUV_{mean,L}}{SUV_{mean,S}} \quad (6)$$

## 3. Results

In order to visually inspect the quality of L-PET images, sagittal views of the entire L-PET images (different low-dose PET images before applying the post-reconstruction filter) together with the S-PET image and the corresponding CT image of a representative patient are shown in Figure 1. The increased levels of signal-to-noise ratio (SNR) are clearly observed with the increased dose levels. The PET image with 1% of the standard dose bears remarkably poor SNR and unclear radiotracer uptake pattern. In addition, Figure 2 illustrates the same views presented in Figure 1 after applying the post-reconstruction Gaussian filters, independently optimized for each dose level.

To quantify the levels of noise in L-PET images, the PSNR, SSIM, and RMSE were measured within the entire head region between 1%, 2%, 3%, 4%, 5%, 6%, 8%, and 10% L-PET images and the reference S-PET images. Table 1 summarizes the average and standard deviation of the above-mentioned metrics calculated across 50 subjects. Table 1 also reports the quantitative metrics measured after applying post-reconstruction smoothing.

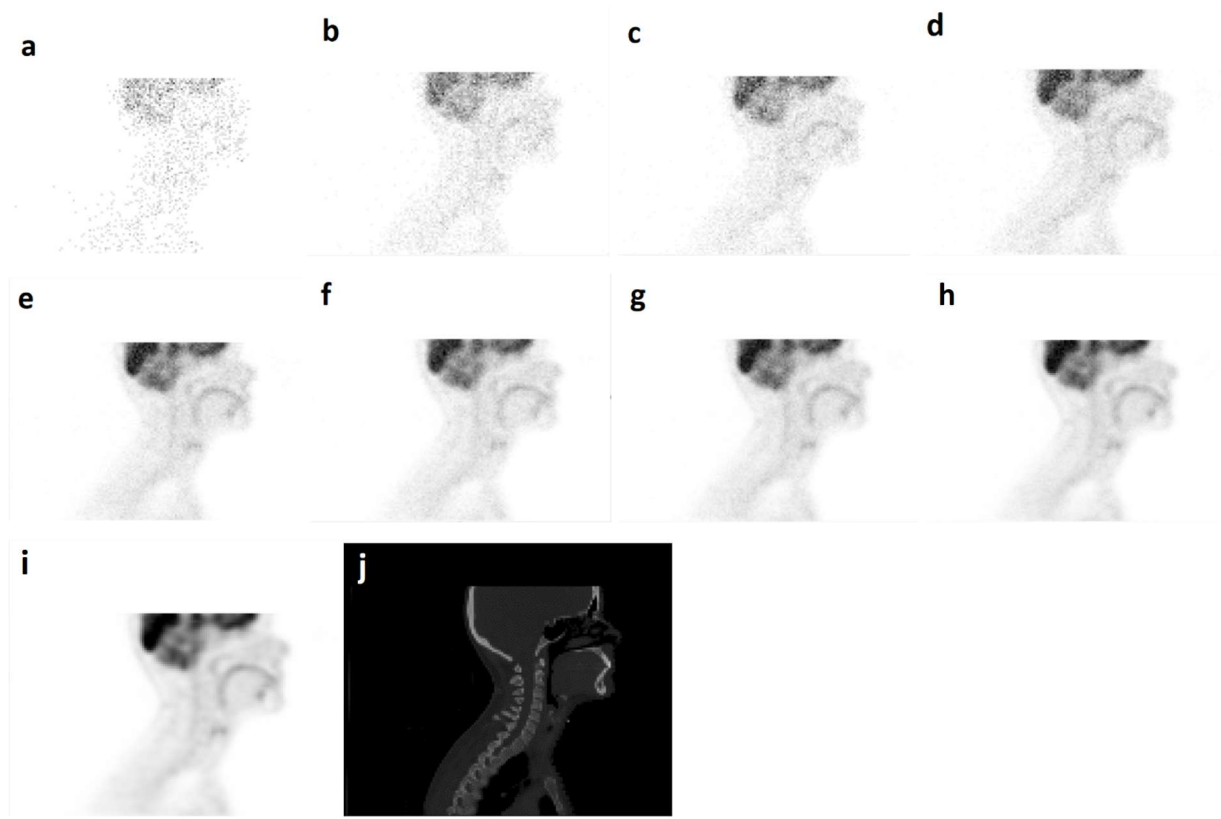

**Figure 1.** Sagittal views of the different low-dose PET images including a) 1%, b) 2%, c) 3%, d) 4%, e) 5%, f) 6%, g) 8%, h) 10%., and i) standard dose PET (100%) image. j) CT image.

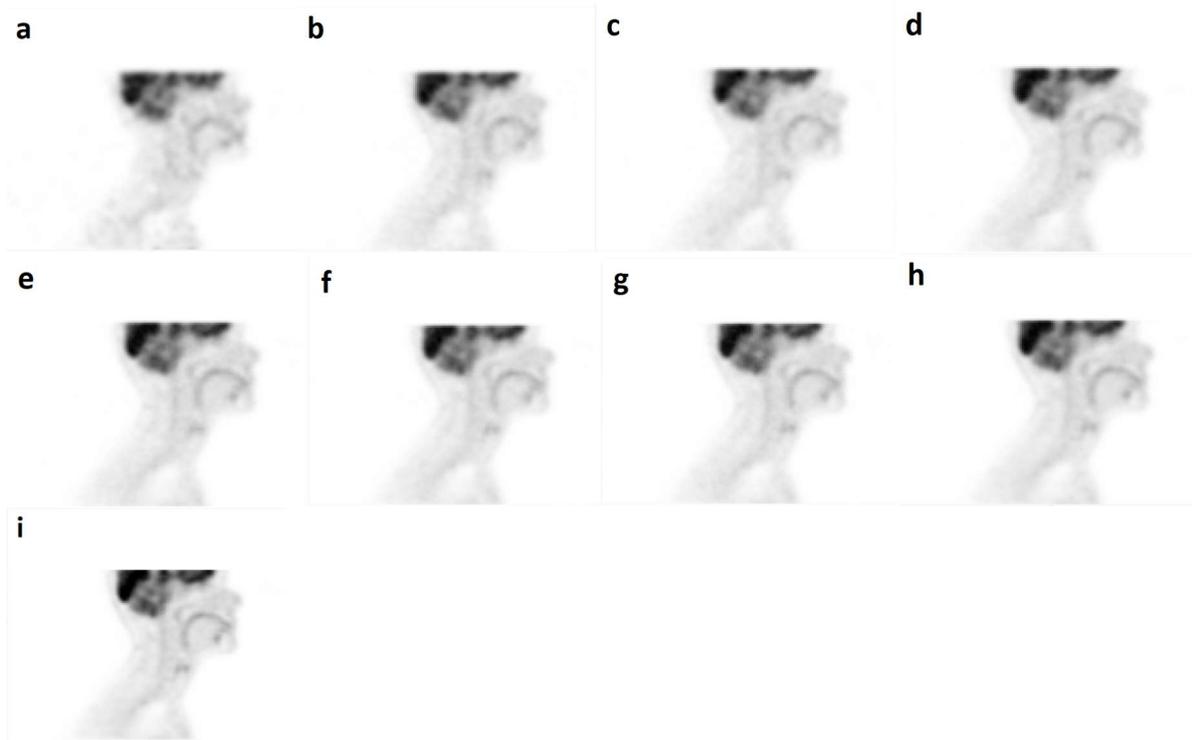

**Figure 2.** Sagittal view of the different low-dose PET images including a) 1%, b) 2%, c) 3%, d) 4%, e) 5%, f) 6%, g) 8%, h) 10% after applying thepost-reconstruction Gaussian filters. i) Standard-dose PET (100%) image.

**Table 1.** The average values and standard deviation of the PSNR, SSIM, and RMSE measured over the entire head region across 50 subjects before and after applying the post-reconstruction Gaussian filters.

| Head | Before post-reconstruction smoothing | | | After post-reconstruction smoothing | | |
|---|---|---|---|---|---|---|
| | PSNR | SSIM | RMSE (SUV) | PSNR | SSIM | RMSE (SUV) |
| 1% | 15.9±1.2 | 0.666±0.061 | 1.454±0.199 | 32.6±2.1 | 0.941±0.013 | 0.215±0.049 |
| 2% | 24.3±1.2 | 0.776±0.047 | 0.549±0.075 | 36.1±2.5 | 0.975±0.006 | 0.147±0.038 |
| 3% | 26.5±1.2 | 0.814±0.041 | 0.426±0.058 | 37.1±2.5 | 0.979±0.005 | 0.130±0.034 |
| 4% | 29.5±1.2 | 0.863±0.033 | 0.301±0.041 | 38.3±2.7 | 0.985±0.004 | 0.114±0.031 |
| 5% | 34.3±1.2 | 0.927±0.019 | 0.173±0.023 | 40.7±2.7 | 0.990±0.003 | 0.086±0.023 |
| 6% | 35.3±1.2 | 0.937±0.016 | 0.155±0.021 | 41.0±2.8 | 0.991±0.002 | 0.083±0.023 |
| 8% | 37.3±1.2 | 0.956±0.012 | 0.123±0.016 | 42.4±2.4 | 0.992±0.002 | 0.071±0.017 |
| 10% | 41.3±1.2 | 0.979±0.005 | 0.077±0.010 | 43.7±3.0 | 0.996±0.001 | 0.062±0.018 |

To meticulously investigate the impact of dose reduction on the quality of PET images, only patients with malignant lesions in the head and neck were included in this study. The quantitative assessment of noise levels before and after applying post-reconstruction smoothing was separately performed within the malignant lesions. Quantification of the malignant lesions in PET images is of special interest in clinical practices for the diagnosis, staging, and treatment monitoring of cancer patients. Figure 3 presents the manifestation of a representative lesion in the different low-dose PET images prior to and following the application of the post-reconstruction filter. Visual inspection revealed significant improvement in tumor detectability after applying the Gaussian filter, in particular for 1% low-dose PET image.

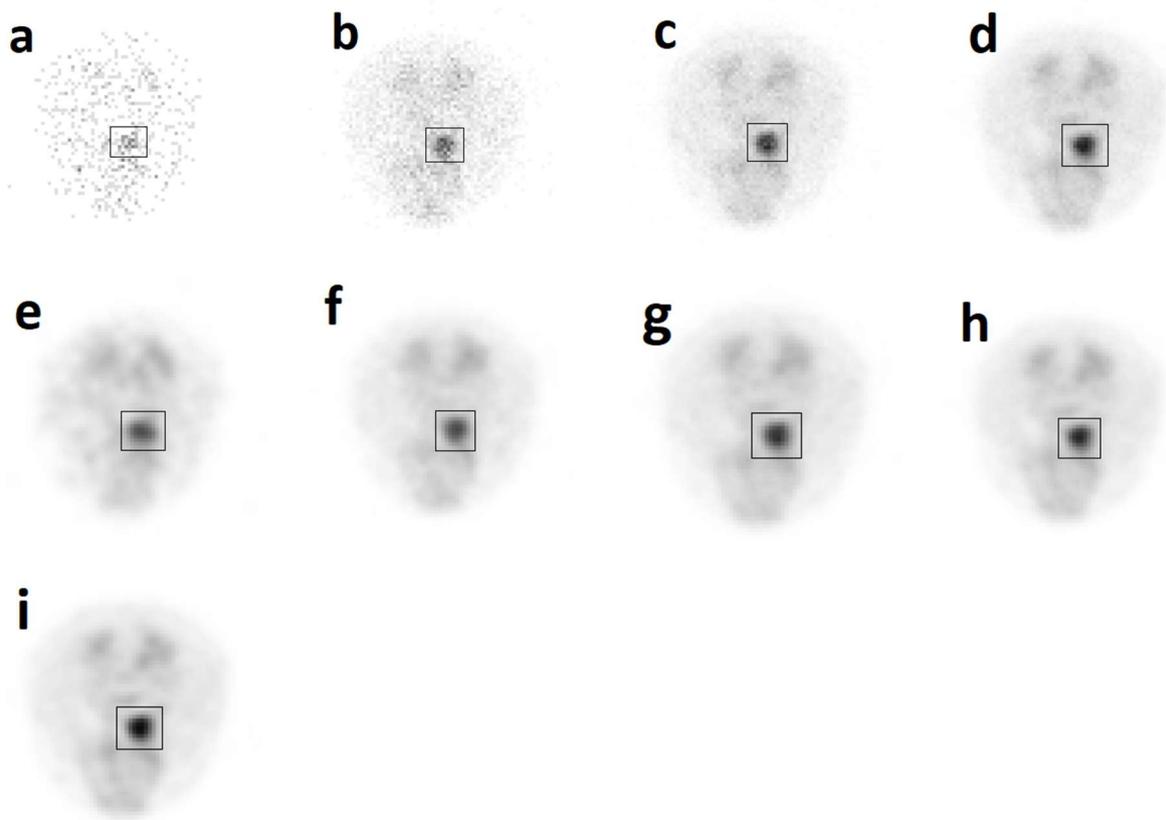

**Figure 3.** Transaxial views of a representative malignant lesion in the different low-dose PET images a) 1%, b) 3%, c) 6%, d) 10% before Gaussian filtering, and e) 1%, f) 3%, g) 6%, h) 10% after Gaussian filtering. i) Standard-dose PET image. The lesions are indicated by the black boxes.

Table 2 summarizes the RMSE, $SUV_{max}$ and $SUV_{mean}$ as well as $SUV_{max}$ and $SUV_{mean}$ relative bias measured within the malignant lesions in the different low-dose PET images before and after applying the Gaussian filter. Noticeable $SUV_{max}$ difference and bias are observed in low-dose PET images (in particular for 1%, 2%, 3% levels) before applying the Gaussian filters.

**Table 2.** The average values and standard deviation of the RMSE, SUV$_{max}$, SUV$_{mean}$ as well as SUV$_{max}$, and SUV$_{mean}$ relative bias measured within the malignant lesions.

| Lesion | Before post-reconstruction smoothing | | | | | After post-reconstruction smoothing | | | | |
|---|---|---|---|---|---|---|---|---|---|---|
| | RMSE (SUV) | SUV$_{max}$ (SUV) | SUV$_{mean}$ (SUV) | SUV$_{max}$ bias (%) | SUV$_{mean}$ bias (%) | RMSE (SUV) | SUV$_{max}$ (SUV) | SUV$_{mean}$ (SUV) | SUV$_{max}$ bias (%) | SUV$_{mean}$ bias (%) |
| 1% | 5.787±1.113 | 36.54±11.241 | 5.496±2.042 | 171.17±16.07 | 0.65±4.40 | 1.191±0.417 | 9.72±4.35 | 5.01±1.92 | -25.47±7.86 | -8.64±3.77 |
| 2% | 2.165±0.420 | 19.65±6.943 | 5.462±2.033 | 52.08±4.89 | 0.02±1.65 | 0.808±0.278 | 10.70±4.65 | 5.16±1.97 | -15.87±6.62 | -5.72±2.29 |
| 3% | 1.693±0.314 | 17.64±5.777 | 5.459±2.021 | 36.45±4.26 | 0.04±1.51 | 0.689±0.254 | 11.01±4.56 | 5.23±1.97 | -13.70±6.20 | -4.45±2.00 |
| 4% | 1.196±0.231 | 15.55±5.380 | 5.460±2.027 | 20.20±5.29 | -0.01±0.93 | 0.594±0.209 | 11.11±4.61 | 5.26±1.98 | -12.21±5.68 | -3.87±1.57 |
| 5% | 0.688±0.130 | 13.91±5.162 | 5.461±2.024 | 10.82±2.00 | 0.01±0.41 | 0.409±0.139 | 11.61±4.64 | 5.33±1.99 | -7.55±4.34 | -2.36±0.83 |
| 6% | 0.616±0.118 | 13.64±4.953 | 5.462±2.019 | 9.08±1.34 | 0.09±0.45 | 0.404±0.145 | 11.58±4.59 | 5.34±1.99 | -7.57±3.83 | -2.30±0.77 |
| 8% | 0.489±0.095 | 13.36±5.113 | 5.461±2.023 | 7.44±2.09 | 0.04±0.42 | 0.325±0.103 | 11.81±4.64 | 5.37±2.01 | -5.36±3.19 | -1.69±0.61 |
| 10% | 0.308±0.056 | 12.87±4.694 | 5.461±2.026 | 2.89±1.90 | -0.004±0.20 | 0.299±0.105 | 11.76±4.57 | 5.37±2.00 | -6.09±3.34 | -1.72±0.54 |
| Standard dose (100%) | ___ | 12.28±4.568 | 5.460±2.025 | ___ | ___ | ___ | 12.28±4.568 | 5.46±2.025 | ___ | ___ |

Moreover, bar plots of SUV$_{mean}$ and SUV$_{max}$ bias for the malignant lesions in the different low-dose PET images before and after post-reconstruction smoothing are presented in Figures 4 and 5, respectively. Remarkable SUV$_{max}$ bias is observed before applying post-reconstruction smoothing, while SUV$_{mean}$ bias raised significantly after applying Gaussian filters.

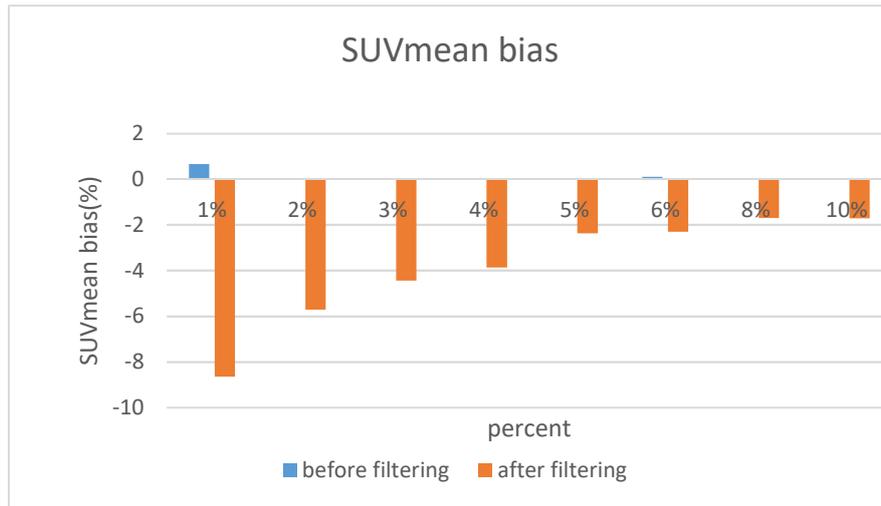

**Figure 4.** Bar plots of the SUV$_{mean}$ bias before and after applying post-reconstruction smoothing for the different noise levels.

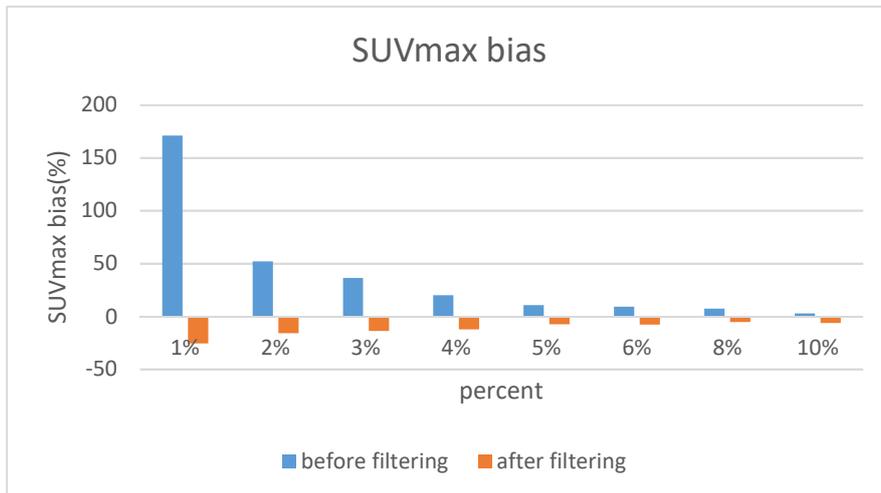

**Figure 5.** Bar plots of the SUV$_{max}$ bias before and after applying post-reconstruction smoothing for the different noise levels.

## 4. Discussion

In this study, low-dose brain PET imaging at different reduced doses of 1%, 2%, 3%, 4%, 5%, 6%, 8%, and 10% of the standard dose was investigated in terms of the overall image quality as well as the quantitative analysis of the activity concentration within the malignant lesions. The standard image quality metrics including PSNR, SSIM, RMSE, and SUV bias were employed to assess the image quality of low-dose PET images before and after applying the post-reconstruction Gaussian filters.

As expected, the PET image quality remarkably improved with an increase in the administrated radiotracer dose, particularly from 1% to 5% of the standard dose, which led to a significant decrease in RMSE from 1.454 SUV to 0.173 SUV. The application of the post-reconstruction Gaussian filters effectively improved the quality of low dose PET images where RMSE decrease to 0.215 SUV and 0.086 SUV for 1% and 5% low-dose PET images, respectively. Regarding the overall quality of the PET images (Table 1), 5% low-dose PET images, after applying the Gaussian filter, exhibited a reasonable compromise between dose reduction and image quality. The post-reconstruction Gaussian filter greatly improved the overall quality of the low-dose PET image resulting in a higher signal-to-noise ratio and signal detectability.

Regarding Table 2, a similar trend was observed in the analysis of malignant lesions wherein RMSEs within the malignant lesions reduced from 5.787 SUV and 0.688 SUV to 1.191 SUV and 0.409 SUV in 1% and 5% low-dose PET images, respectively, after applying the Gaussian filter. Since the mean and max SUV within the malignant lesions are commonly used in clinical practices for the assessment/diagnosis/staging of the tumors; these metrics were separately investigated in low-dose PET images. Figure 5 shows that SUV max within malignant lesions enormously changes in low-dose PET images; however, after applying the Gaussian filter, the SUV max bias significantly reduced. It is strongly recommended that the SUV max in low-dose PET images should be considered after applying the post-reconstruction filter to avoid noise-induced bias.

On the other hand, the analysis of SUV mean demonstrated that the low-dose PET images before applying the post-reconstruction filter bear insignificant mean SUV bias within the malignant lesions (Figure 4). However, after applying the Gaussian filter, this mean SUV bias increased significantly (up to 8%), which might skew the accurate diagnosis. In clinical practices, mean SUV is commonly used to assess malignant lesions; in this light, the mean SUV should be considered before applying the post-reconstruction filters to avoid the bias imposed by the smoothing filter.

## 5. Conclusion

In this work, low-dose PET imaging was systematically investigated at different percentages of the standard injection dose. Overall, 5% low-dose PET images after applying the post-reconstruction Gaussian filter exhibited a reasonable compromise between reducing the injected activity and image quality. The quantitative analysis of the malignant lesion demonstrated that mean SUV bias in different low-dose PET images is negligible before the application of the post-reconstruction filter, while significant mean SUV bias was observed after applying the Gaussian filter. On the other hand, max SUV bias was noticeably reduced after applying the Gaussian filter. It is strongly recommended that the mean SUV should be considered before applying the post-reconstruction filter, and max SUV should be considered after applying the post-reconstruction filter in low-dose PET imaging.

**References**

[1] Basu S, Hess S, Nielsen Braad PE, Olsen BB, Inglev S, Høilund-Carlsen PF. The Basic Principles of FDG-PET/CT Imaging. PET clinics. 2014;9:355-70, v.
[2] Zimmer L. [PET imaging for better understanding of normal and pathological neurotransmission]. Biologie aujourd'hui. 2019;213:109-20.
[3] Arabi H, Zeraatkar N, Ay MR, Zaidi H. Quantitative assessment of inter-crystal scatter and penetration in the PET subsystem of the FLEX triumph preclinical multi-modality scanner. Iranian Journal of Nuclear Medicine. 2010;18:40.
[4] Lameka K, Farwell MD, Ichise M. Positron Emission Tomography. Handbook of clinical neurology. 2016;135:209-27.
[5] Arabi H, Zaidi H. Non-local mean denoising using multiple PET reconstructions. Annals of nuclear medicine. 2020.
[6] Arabi H, Zaidi H. Deep learning-guided estimation of attenuation correction factors from time-of-flight PET emission data. Medical image analysis. 2020;64:101718.
[7] Li Y, Jiang L, Wang H, Cai H, Xiang Y, Li L. EFFECTIVE RADIATION DOSE OF 18F-FDG PET/CT: HOW MUCH DOES DIAGNOSTIC CT CONTRIBUTE? Radiation protection dosimetry. 2019;187:183-90.
[8] Khoshyari-morad Z, Jahangir R, Miri-Hakimabad H, Mohammadi N, Arabi H. Monte Carlo-based estimation of patient absorbed dose in 99mTc-DMSA, -MAG3, and -DTPA SPECT imaging using the University of Florida (UF) phantoms. arXiv:210300619. 2021.
[9] Mehranian A, Arabi H, Zaidi H. Quantitative analysis of MRI-guided attenuation correction techniques in time-of-flight brain PET/MRI. NeuroImage. 2016;130:123-33.
[10] Arabi H, Zaidi H. Improvement of image quality in PET using post-reconstruction hybrid spatial-frequency domain filtering. Physics in medicine and biology. 2018;63:215010.
[11] Arabi H, Zaidi H. Spatially-guided non-local mean filter for denoising of clinical whole-body PET images.  2018 IEEE Nuclear Science Symposium and Medical Imaging Conference Proceedings (NSS/MIC)2018. p. 1-3.
[12] Zhou L, Schaefferkoetter JD, Tham IWK, Huang G, Yan J. Supervised learning with cyclegan for low-dose FDG PET image denoising. Medical image analysis. 2020;65:101770.
[13] Schaefferkoetter JD, Yan J, Townsend DW, Conti M. Initial assessment of image quality for low-dose PET: evaluation of lesion detectability. Physics in medicine and biology. 2015;60:5543-56.
[14] Yan J, Schaefferkoette J, Conti M, Townsend D. A method to assess image quality for Low-dose PET: analysis of SNR, CNR, bias and image noise. Cancer imaging : the official publication of the International Cancer Imaging Society. 2016;16:26.
[15] Oehmigen M, Ziegler S, Jakoby BW, Georgi JC, Paulus DH, Quick HH. Radiotracer dose reduction in integrated PET/MR: implications from national electrical manufacturers association phantom studies. Journal of nuclear medicine : official publication, Society of Nuclear Medicine. 2014;55:1361-7.
[16] Arabi H, Zaidi H. Applications of artificial intelligence and deep learning in molecular imaging and radiotherapy. Euro J Hybrid Imaging. 2020;4:17.
[17] Sanaat A, Arabi H, Mainta I, Garibotto V, Zaidi H. Projection Space Implementation of Deep Learning-Guided Low-Dose Brain PET Imaging Improves Performance over Implementation in Image Space. Journal of nuclear medicine : official publication, Society of Nuclear Medicine. 2020;61:1388-96.
[18] Arabi H, Zaidi H. Dual domain spatial-transform smoothing of whole-body PET images. Journal of Nuclear Medicine. 2017;58:616-.
[19] Arabi H, Zaidi H. Spatially guided nonlocal mean approach for denoising of PET images. Medical physics. 2020;47:1656-69.

[20] Ghane B, Karimian A, Mostafapour S, Gholamiankhak F, Shojaerazavi S, Arabi H. Quantitative analysis of image quality in low-dose CT imaging for Covid-19 patients. arXiv preprint arXiv:210208128. 2021.